\input amstex
\documentstyle{amsppt}
\magnification=\magstep1                        
\hsize6.5truein\vsize8.9truein                  
\NoRunningHeads
\loadeusm

\magnification=\magstep1                        
\hsize6.5truein\vsize8.9truein                  
\NoRunningHeads
\loadeusm

\document
\topmatter

\title
Riesz Polarization Inequalities in Higher Dimensions
\endtitle

\rightheadtext{the multiplicity of the zero at $1$ of polynomials}

\author
Tam\'as Erd\'elyi and Edward B. Saff*
\endauthor

\address Department of Mathematics, Texas A\&M University,
College Station, Texas 77843 (T. Erd\'elyi)
\endaddress

\address Center for Constructive Approximation, 1326 Stevenson Center, Vanderbilt University,
Nashville, TN 37240 (E.B. Saff)
\endaddress
\email terdelyi\@math.tamu.edu,  edward.b.saff\@vanderbilt.edu
\endemail

\endthanks

\endkeywords

\thanks {*The research of this author was supported, in part, by U.S. National Science
Foundation grants DMS-0808093 and DMS-1109266.}\endthanks

\abstract
We derive bounds and asymptotics for the maximum Riesz polarization quantity
$$M_n^p(A) := \max_{{\bold x}_1, {\bold x}_2, \ldots, {\bold x}_n \in A}
{\min_{{\bold x} \in A}{\sum_{j=1}^n{\frac{1}{|{\bold x} - {\bold x}_j|^{p}}}}}$$
(which is  $n$ times the Chebyshev constant ) for quite general sets $A \subset {\Bbb R}^m$ with special focus on the unit sphere and
unit ball. We combine elementary averaging arguments with potential theoretic tools to
formulate and prove our results. We also give a discrete version of the recent result of Hardin, Kendall,
 and Saff which solves the Riesz polarization problem for the case when $A$ is the unit circle and $p>0,$ as well
 as provide an independent proof of their result for $p=4$ that exploits classical polynomial inequalities
 and yields new estimates. Furthermore, we raise some challenging conjectures.
\endabstract

\endtopmatter

\document

\head {1. Introduction} \endhead

For $n \in {\Bbb N}$, let $\omega_n=\{{\bold x}_1,{\bold x}_2, \ldots, {\bold x}_n\}$
denote $n$ (not necessarily distinct) points in $m$-dimensional Euclidean space
${\Bbb R}^{m}.$  We define for  $p > 0$ and a compact set $A \subset {\Bbb R}^m$, the
{\it Riesz polarization quantities}
$$ M^p(\omega_n,A) := \min_{{\bold x}\in A}
\sum_{j=1}^n \frac{1}{|{\bold x} - {\bold x}_j|^{p}}\,,\quad M_n^p(A):=
\max_{\omega_n \subset A} M^p(\omega_n,A).\tag 1.1$$
Such max-min quantities for potentials were first introduced by M. Ohtsuka who
explored (for very general kernels) their relationship to various definitions of capacity that arise in electrostatics (see [O-67]). In particular,
he showed that for any compact set $A \subset {\Bbb R}^m$ the following limit, called the {\it Chebyshev constant} of $A$, exists as an extended
real number:
$${\Cal M}^p(A):= \lim_{n \to \infty}\frac{M_n^p(A)}{n}. \tag 1.2$$
Moreover, he showed that ${\Cal M}^p(A)$ is not smaller than the Wiener constant $W_p(A)$ for $A$ (see Section 2). In this
paper we primarily focus on results when  the set $A$ is the unit sphere or the unit ball and consider both the cases when the limit (1.2) is
finite and when it is infinite.

In his Ph.D. dissertation [A-09], G. Ambrus proved the following basic result for the case
when $A \subset {\Bbb R}^{2} $ is the unit circle ${\Bbb S}^1$ and $p=2.$

\proclaim{Theorem 1.1}
We have
$$M_n^2({\Bbb S}^1) = \frac{n^2}{4}\,, \qquad n \geq 1\,,\tag 1.3$$
and $M^2(\omega_n,{\Bbb S}^1)=n^2/4, \,\omega_n \subset {\Bbb S}^1,$  if and only if the $n$ points of
$\omega_n$ are equally spaced on ${\Bbb S}^1.$
\endproclaim

In [ABE-12], Ambrus's rather technical proof along with a simpler proof based on Bernstein's
inequality for entire functions are presented.  Bernstein's inequality
was also used in [ABE-12] to provide an equally simple proof of the following estimates for the
unit circle.

\proclaim{Theorem 1.2}
For $n \geq 2$ we have
$$M_n^p({\Bbb S}^1)  \leq
\cases c_pn^p \,, \quad &  \enskip p > 1\,, \\
c_1n\log n\,, \quad &  \enskip p = 1\,, \\
\displaystyle{\frac{c_0n}{1-p}}\,, \quad & \enskip p \in [0,1)\,,
\endcases$$
for some constants $c_p > 0$ depending only on $p \geq 1$ and an absolute constant $c_0 > 0$.
\endproclaim

In Section 2 we use minimum energy methods and potential theory to obtain estimates for
$M_n^p(A)$ for a large class of sets $A \subset {\Bbb R}^{m}.$
In Section 3 we apply the results of Section 2 to obtain  higher dimensional analogs of
Theorem 1.2 for the unit sphere as well as for the unit ball.

In Section 4 we return to the case of the unit circle of the complex plane.
For all $p>0,$ it is conjectured in [ABE-12] that the maximum polarization on ${\Bbb S}^1$ occurs for
the $n$-th roots of unity $\omega_n^*:=\{ e^{i2\pi k/n}\,: k=1,2,\ldots,n\};$ that is,
$$M_n^p({\Bbb S}^1)=M^p(\omega_n^*,{\Bbb S}^1).\tag 1.4$$
This conjecture was recently proved by Hardin, Kendall, and Saff 
in [HKS-12]. Here, we provide some additional consequences of their argument. Furthermore, by exploring connections to 
classical polynomial
inequalities, we provide an independent proof of the conjecture for $p=4$, namely that
$$M_n^4({\Bbb S}^1) = \frac{n^4}{48} + \frac{n^2}{24}\,,\tag 1.5$$
where the maximum is attained for $n$ distinct equally spaced points on the unit circle.
Although our argument (obtained prior to the general result in [HKS-12]) is not brief, it does yield additional inequalities for the discrete Riesz potential  in this special case.

In Section 5, we provide the proofs of results stated in Sections 2 and 3.

We call the reader's attention to two recent articles [NR-12a] and
[NR-12b] that contain somewhat related results for the extrema of sums of certain
powered distances to finite point sets.

\head 2. Polarization inequalities via energy methods \endhead

For a set $\omega_n = \{{\bold x}_1, {\bold x}_2, \ldots, {\bold x}_n\}$ of $n(\geq 2)$
distinct points in ${\Bbb R}^m$, we define the {\it Riesz $p$-energy} of $\omega_n$ by
$$E_p(\omega_n) := \sum_{j \neq k}{\frac{1}{|{\bold x}_j - {\bold x}_k|^p}} =
2\sum_{1 \leq j < k \leq n}{\frac{1}{|{\bold x}_j - {\bold x}_k|^p}}\,,$$
and we consider the {\it minimum $n$-point Riesz $p$-energy} of an infinite compact set
$A \subset {\Bbb R}^m$ defined by
$${\Cal E}_p(A;n) := \min \{E_p(\omega_n): \omega_n \subset A,\,|\omega_n|=n\}\,. \tag 2.1$$
We denote by $\omega_{n,p}^* = \{{\bold x}_1^*, {\bold x}_2^*, \ldots , {\bold x}_n^*\}$
an $n$-point $p$-energy minimizing configuration on $A$; i.e.,
$E_p(\omega_{n,p}^*) = {\Cal E}_p(A;n)$.
Further we denote by $U_{n,p}^*({\bold x})$ the potential function associated with
$\omega_{n,p}^*$; i.e.,
$$U_{n,p}^*({\bold x}) := \sum_{j=1}^n{|{\bold x} - {\bold x}_j^*|^{-p}}\,.$$
It is well-known (and easy to show) that
$$(n-1){\Cal E}_p(A;n+1) \geq (n+1) {\Cal E}_p(A;n)\,, \tag 2.2$$
from which it follows that
$$C^*(A,n,p) := \min\{U_{n,p}^*({\bold x}): {\bold x} \in A\} \geq
\frac{1}{n-1} \, {\Cal E}_p(A;n)\,; \tag 2.3$$
indeed, we have
$$2C^*(A,n,p) + {\Cal E}_p(A;n) \geq {\Cal E}_p(A;n+1)\,,$$
and after multiplying this inequality by $n-1$ and applying (2.2), we get (2.3). Thus lower
estimates for ${\Cal E}_p(A;n)$ yield lower estimates for $M_n^p(A).$

We next mention some known asymptotic results for ${\Cal E}_p(A;n)$ as $n \rightarrow \infty$.
The following theorem appearing in [HS-05] and [BHS-08] has been referred to as the
{\it Poppy-seed Bagel Theorem} because of its interpretation for distributing points on a
torus.

\proclaim{Theorem 2.1}
Let $d \in {\Bbb N}$ and $A \subset {\Bbb R}^m$ be an infinite compact $d$-rectifiable set.
Then for $p > d$ we have
$$\lim_{n \rightarrow \infty} {\frac {{\Cal E}_p(A;n)}{n^{1+p/d}}} =
\frac{C_{p,d}}{{\Cal H}_d(A)^{p/d}}\,, \tag 2.4$$
where $C_{p,d}$ is a finite positive constant (independent of $A$ and $m$) and
${\Cal H}_d(\cdot)$ denotes the $d$-dimensional Hausdorff measure in ${\Bbb R}^m$ normalized
so that an embedded $d$-dimensional unit cube has measure $1$.
\endproclaim

By a {\it $d$-rectifiable set} we mean the Lipschitz image of a bounded set in ${\Bbb R}^d$.

In [MRS-04, Theorem 3.1] it is shown that $C_{p,1}$ can be expressed in terms of the classical
 Riemann zeta function; namely $C_{p,1}= 2 \zeta(p)$. For $d \geq 2$ the precise value of
$C_{p,d}$ is not known. The significance (and difficulty) of determining $C_{p,d}$ is deeply
rooted in its  connection to densest sphere packings in ${\Bbb R}^d$. For $d=2$ it is
conjectured in [KS-98] that $C_{p,2}=(\sqrt{3}/2)^{p/2}\zeta_{L}(p),$ where $L$ denotes the
planar hexagonal lattice of points $m(1,0)+n(1/2,\sqrt{3}/2),\, m,n \in {\Bbb Z},$ and
$\zeta_{L}$ is the Epstein zeta function $\zeta_{L}(p):=\sum _{X\in L,X\neq 0}|X|^{-p}.$

Concerning lower estimates for $C_{p,d}$, it follows from [BHS-12, Proposition 4] that, for
$p > d \geq 2$ and $\frac 12 (p-d)$ not an integer,
$$C_{p,d} \geq \frac{d\pi^{p/2}}{p-d}
\left( \frac{\Gamma(1+\frac{p-d}{2})}{\Gamma(1+\frac{p}{2})} \right)^{p/d}\,. \tag 2.5$$
For the case $p=d$, the minimum $p$-energy grows like $n^2\log n$. The following result is
given in [HS-05].

\proclaim{Theorem 2.2}
Let $d \in {\Bbb N}$ and $A$ be an infinite compact subset of a $d$-dimensional
$C^1$-manifold embedded in ${\Bbb R}^m$. Then
$$\lim_{n \rightarrow \infty} {\frac {{\Cal E}_d(A;n)}{n^2 \log n}} =
\frac {\beta_d}{{\Cal H}_d(A)}\,,$$
where $\beta_d$ is the volume of the $d$-dimensional unit ball.
\endproclaim

For the case when $0 < p < d :=$dim$(A)$, the Hausdorff dimension of $A$, a theorem from
classical potential theory (cf., e.g. [L-72]) asserts that
$$\lim_{n \rightarrow \infty}{\frac{{\Cal E}_p(A;n)}{n^2}} = W_p(A)\,, \tag 2.6$$
where $W_p(A)$ is the so-called {\it Wiener constant} defined by
$$W_p(A) :=
\inf \iint{\frac {1}{|{\bold x} - {\bold y}|^p} \, d\mu({\bold x}) \, d\mu({\bold y})}\,,$$
the infimum being taken over all Borel probability measures $\mu$ supported on $A$.

From the above results and observations we immediately obtain

\proclaim{Theorem 2.3}
If $A \subset {\Bbb R}^m$ is an infinite compact set, then
$$M_n^p(A) \geq \frac{1}{n-1} \, {\Cal E}_p(A;n)\,, \qquad n \geq 2\,.\tag 2.7$$
Let $d \in {\Bbb N}$. If $A$ is $d$-rectifiable, then
$$\liminf_{n \rightarrow \infty}{\frac{M_n^p(A)}{n^{p/d}}} \geq
\frac{C_{p,d}}{\left({\Cal H}_d(A)\right)^{p/d}}\,, \qquad p > d\,, \tag 2.8$$
where the constant $C_{p,d}$ is given in Theorem 2.1.\

If $A$ is any infinite compact subset of a $d$-dimensional $C^1$-manifold, then
$$\liminf_{n \rightarrow \infty}{\frac{M_n^d(A)}{n\log n}} \geq
\frac{\beta_d}{{\Cal H}_d(A)}\,, \qquad p=d.\, \tag 2.9$$\

If $A$ is any infinite compact subset of ${\Bbb R}^m,$ then
$${\Cal M}^p(A)=\lim_{n \rightarrow \infty}{\frac{M_n^p(A)}{n}} \geq W_p(A)\,,
\qquad 0 < p < d = \text {\rm dim}(A)\,. \tag 2.10$$
\endproclaim
We remark that inequality (1.7)  appears in [FN-08] and [FR-06]. Also, as previously mentioned, the
inequality (2.10) is proved in [O-67]. Moreover, it follows from [FN-08, Theorem 11] that equality holds in
(2.10) whenever the maximum principle is satisfied on $A$ for Riesz potentials having kernel $K({\bold x},{\bold y})=|{\bold x}-{\bold y}|^{-p}$.

Regarding upper bounds for $M_n^p(A),$ standard arguments (see Section 5) yield the following.

\proclaim{Theorem 2.4} Let $A \subset {\Bbb R}^m$ be an infinite compact set. If ${\Cal H}_d(A)>0,$ then there exists a constant $c_p > 0$ depending
only on $p$ such that
$$M_n^p(A)\leq \displaystyle {\frac{c_p}{p-d}\,n^{p/d}},
\qquad p>d\,, \enskip n \geq 1\,, \tag 2.11$$
and there exists an absolute constant $c_1 > 0$ such that
$$M_n^d(A)\leq c_1n\log n\,, \qquad p=d\,, \enskip n \geq 2\,. \tag 2.12$$
If there exists
a probability measure $\mu_A$ supported on $A$ whose $p$-potential is bounded on $A,$ say
$$\int{\frac {1}{|{\bold x} - {\bold y}|^p} \, d\mu_A({\bold y})} \leq w_p,
\qquad {\bold x} \in A\,,$$
then
$$M_n^p(A) \leq nw_p\,, \qquad p > 0\,, \enskip n \geq 1\,. \tag 2.13$$
\endproclaim

The essential property used in the proof of Theorem 2.4 given in Section 5 is that $A$ is upper $d$-regular with respect to a Borel probability measure
$\mu$ supported on $A;$ that is, there exists a positive constant $C_0$ such that for
any open ball $B^m({\bold x},r)\subset {\Bbb R}^m$ with
center ${\bold x} \in A$ and radius $r>0$ there holds
$$\mu(B^m({\bold x},r)\cap A) \leq C_0 r^d. \tag 2.14$$
This property is a consequence of Frostman's Lemma (see [M-95, Chap. 8]).

\head 3. Polarization Inequalities for the Unit Sphere and Unit Ball \endhead

Let
$${\Bbb S}^d := \{{\bold x} \in {\Bbb R}^{d+1}: |{\bold x}| = 1\} \quad \text {and} \quad
{\Bbb B}^d  := \{{\bold x} \in {\Bbb R}^d:|{\bold x}| \leq 1\}\,.\tag 3.1$$
Utilizing the results of Section 2 together with the known facts (cf. [L-72]) that
$$\split W_p({\Bbb S}^d) = & \iint
{\frac {1}{|{\bold x} - {\bold y}|^p} \, d\sigma_d({\bold x}) \, d\sigma_d({\bold y})} \cr
= & \, 2^{d-p-1} \, \frac{\Gamma \big(\frac{d+1}{2}\big) \Gamma \big(\frac{d-p}{2}\big)}
{\sqrt{\pi} \, \Gamma \big(d-\frac{p}{2}\big)}\,, \qquad 0 < p < d\,, \cr \endsplit \tag 3.2$$
where $\sigma_d$ denotes the normalized surface area on ${\Bbb S}^d$, and
$$W_p({\Bbb B}^d) =
\frac{\Gamma\big(\frac{d-p}{2}\big) \, \Gamma \big(\frac p2 + 1\big)}
{\Gamma\big( \frac d2 \big)}\,,
\qquad d-2 \leq p < d\,, \enskip p > 0\,, \tag 3.3$$
we shall prove the following two theorems.

\proclaim{Theorem 3.1} For the sphere ${\Bbb S}^d, \,d \geq 2,$ we have
$$\liminf_{n \rightarrow \infty}{\frac{M_n^p({\Bbb S}^d)}{n^{p/d}}} \geq
C_{p,d} \, \left(\frac{\Gamma\big(\frac{d+1}{2}\big)}{2\pi^{(d+1)/2}}\right)^{p/d}\,,
\qquad p > d\,; \tag 3.4$$
$$\lim_{n \rightarrow \infty}{\frac{M_n^p({\Bbb S}^d)}{n\log n}} =
\frac 1d \frac{\Gamma \big(\frac{d+1}{2} \big)}{\sqrt{\pi} \, \Gamma \big(\frac d2 \big)}
=:\tau_d\,, \qquad p=d\,; \tag 3.5$$
$$\lim_{n \rightarrow \infty}{\frac{M_n^p({\Bbb S}^d)}{n}}=
2^{d-p-1} \, \frac{\Gamma\big(\frac{d+1}{2} \big)\Gamma \big(\frac{d-p}{2} \big)}
{\sqrt{\pi} \, \Gamma \big( d-\frac p2 \big)}\,, \qquad 0 < p < d\,. \tag 3.6$$
Furthermore, the following upper estimates hold for all $n \geq 3$.
$$ M_n^p({\Bbb S}^d) \leq
\cases \displaystyle{\left( \frac {np\tau_d}{p-d} \right)^{p/d}}\,, \qquad & p>d\,, \\
\displaystyle{\tau_d\frac{n[\log n +\log(\log n)+\log(2^d\tau_d) ]}{1-(\log n)^{-1}}}\,,
\qquad & p=d\,, \\
\displaystyle{n 2^{d-p-1} \frac{\Gamma \big( \frac{d+1}{2} \big)
\Gamma \big( \frac{d-p}{2} \big)}{\sqrt{\pi}\,\Gamma \big(d-\frac p2 \big)}}\,,
\qquad & 0<p<d\,.
\endcases \tag 3.7$$
\endproclaim

\proclaim{Theorem 3.2} For the unit ball ${\Bbb B}^d,$ we have
$$\liminf_{n \rightarrow \infty}{\frac{M_n^p({\Bbb B}^d)}{n^{p/d}}} \geq
C_{p,d} \, \left(\frac{\Gamma \big(\frac d2 +1\big)}{\pi^{d/2}}\right)^{p/d}\,,
\qquad p > d\,; \tag 3.8$$
$$\lim_{n \rightarrow \infty}{\frac{M_n^p({\Bbb B}^d)}{n\log n}} = 1\,,
\qquad p=d\,; \tag 3.9$$
$$\frac{M_n^p({\Bbb B}^d)}{n} =1, \qquad 0 < p \leq d-2, \,\,\,n \geq 1; \tag 3.10$$
$$\lim_{n \rightarrow \infty}{\frac{M_n^p({\Bbb B}^d)}{n}} =
\frac{\Gamma \big(\frac {d-p}{2} \big) \Gamma \big(\frac p2 + 1 \big)}
{\Gamma \big(\frac d2 \big)}\,, \qquad 0 \leq d-2 < p < d\,, \enskip p > 0\,. \tag 3.11$$
Furthermore, the following upper estimates hold for all $n \geq 3$:
$$M_n^p({\Bbb B}^d) \leq
\cases \displaystyle{\left( \frac {pn}{p-d} \right)^{p/d}}\,, \qquad & p>d\,, \\
\displaystyle{\frac{n[\log n + \log (\log n) + d\log 2]}{1 - (\log n)^{-1}}}\,,
\qquad & p=d\,, \\
\displaystyle{\frac{n\Gamma \big(\frac{d-p}{2} \big)\Gamma \big(\frac p2 +1 \big)}
{\Gamma \big(\frac d2 \big)}}\,  \qquad & d-2 < p < d\,, \enskip p > 0\,.
\endcases \tag 3.12$$
\endproclaim

\noindent ${\bold {Remark\, 1.}}$ It is easily seen that for  $p > d$ and $n \geq 2^d$, we have
$M_n^p({\Bbb B}^d) \geq 4^{-p}n^{p/d}.$ Indeed, let
$\{{\bold x}_1, {\bold x}_2,\ldots, {\bold x}_m\}$ be a maximal $\delta$-net in
${\Bbb B}^d$ with $\delta := 4n^{-1/d}$. Then
$$m\beta_d(\delta/2)^d \leq \beta_d(1+\delta/2)^d,$$
so
$$m \leq \left( \frac{1+\delta/2}{\delta/2} \right)^d
\leq \left( \frac{4}{\delta} \right)^d \leq n\,.$$
Also, for every ${\bold x} \in {\Bbb B}^d$, there is an
${\bold x}_k \in \{{\bold x}_1, {\bold x}_2,\ldots, {\bold x}_m\}$
such that $|{\bold x} - {\bold x}_k| \leq \delta$. Therefore,
$$\sum_{j=1}^m{|{\bold x} - {\bold x}_j|^{-p}} \geq |{\bold x} - {\bold x}_k|^{-p}
\geq \delta^{-p} = 4^{-p}n^{p/d}\,.$$

Observe further that for the case $0<p<d,$ we have $M_n^p({\Bbb B}^d) \geq n$
since we can take all the points ${\bold x}_j$ equal to ${\bold 0}$, the center of the
unit ball ${\Bbb B}^d,$\ and, moreover, such points are optimal in the case
when $0<p\leq d-2$ (see the proof of (3.10) in Section 5).

\noindent ${\bold{Remark\, 2.}}$ For the case $p>d$ the above theorems establish the
asymptotically sharp order (namely $n^{p/d}$ ) but not the sharp coefficient for the unit
sphere and unit ball. Note, however, from the lower estimates in (2.5), (3.4) and  (3.8)
that, for $A = {\Bbb B}^d$ or $A = {\Bbb S}^d,$ we have
$$\lim_{p \rightarrow d^+}
{\left( \liminf_{n \rightarrow \infty}{\frac{M_n^p(A)}{n^{p/d}}} \right)} = \infty\,.$$
This is clearly consistent with the upper bounds provided in Theorems 3.1 and 3.2 for the
case $p>d.$\

We conclude this section with the following conjectures, which would be an analogs of
Theorems 2.1 and 2.2.
\proclaim{Conjecture 1}
Let $p > d$ and $m\geq d$, where $p$ and $m$ are integers. For every
infinite compact $d$-rectifiable set $A$ in ${\Bbb R}^{m}$, we have
$$
\lim_{n\to \infty}\frac {M^{p}_n(A)}{n^{p/d}}=\frac {\sigma_{p,d}}{{\Cal H}_d(A)^{p/d}}, \tag 3.13
$$
where $\sigma_{p,d}$ is a positive and finite constant independent of $A$ and $m$.

Moreover, if $A$ is $d$-rectifiable with ${\Cal H}_d(A) > 0$, then any sequence $\{\omega_n^\ast\}_{n=2}^{\infty}$ of $p$-polarization maximizing configurations on $A$   is asymptotically uniformly
distributed on $A$ with respect to ${\Cal H}_d$.
\endproclaim
In particular, (1.4)  implies that the constant $\sigma_{p,1}$ appearing in this conjecture would have to  equal   $2(2^p-1)\zeta(p).$

\proclaim{Conjecture 2}
Let $d \in {\Bbb N}$ and $A$ be an infinite compact subset of a $d$-dimensional
$C^1$-manifold embedded in ${\Bbb R}^m$. Then
$$\lim_{n \rightarrow \infty} {\frac {M^{p}_n(A)}{n \log n}} =
\frac {\beta_d}{{\Cal H}_d(A)}\,, \tag 3.14$$
where $\beta_d$ is the volume of the $d$-dimensional unit ball.
   \endproclaim
The results of this section assert that (3.14) holds for spheres and balls.


\head 4. Polarization  on the unit circle \endhead

In this section we explore some connections between polynomial inequalities and the polarization inequality recently proved in [HKS-12].
Let $g$ be a positive-valued even function defined on ${\Bbb R} \setminus (2\pi {\Bbb Z})$ that is periodic
with period $2\pi$. We denote by $\Omega_n$  the collection of all sets
$$\omega_n := \{t_1 < t_2 < \cdots < t_n\} \subset [0,2\pi)\,$$
and put
$$\widetilde{\omega}_n := \{\widetilde{t}_1 < \widetilde{t}_2 < \cdots < \widetilde{t}_n\} \subset [0,2\pi)$$
with
$$\widetilde{t}_j := 2(j-1)\pi/n\,, \qquad j=1,2,\ldots,n\,.$$
We introduce the notation
$$P_{\omega_n(t)} := \sum_{j=1}^n{g(t-t_j)},\quad \quad P_{\widetilde{\omega}_n}(t) := \sum_{j=1}^n{g(t-\widetilde{t}_j)}\,. $$

In [HKS-12] the following theorem is proved.

\proclaim{Theorem 4.1}
Let $g$ be a positive-valued even function defined on ${\Bbb R} \setminus (2\pi {\Bbb Z})$ that is periodic
with period $2\pi$. Suppose that $g$ is non-increasing and strictly convex on $(0,\pi]$. Let $\omega_n \subset [0,2\pi)$.
We have
$$\max_{\omega_n \in \Omega_n}\left\{ \min_{t \in [-\pi,\pi)}{P_{\omega_n}(t)} \right\} = P_{\widetilde{\omega}_n}(\pi/n)\,.$$
\endproclaim

In fact, a closer look at the proof of the main result in [HKS-12] shows that the following
Riesz lemma type improvement also holds.

\proclaim{Theorem 4.2}
Let $g$ be a positive-valued even function defined on ${\Bbb R} \setminus (2\pi {\Bbb Z})$ that is periodic
with period $2\pi$. Suppose that $g$ is non-increasing and strictly convex on $(0,\pi]$. Let $\omega_n \subset [0,2\pi)$.
There is a number $\gamma \in [0,2\pi)$ (depending on $\omega_n$) such that
$$P_{\omega_n}(t) \leq P_{\widetilde{\omega}_n}(t-\gamma)\,, \qquad t \in (\gamma,\gamma + 2\pi/n)\,,$$
for every $\omega_n \in \Omega_n$.
\endproclaim

A consequence of Theorem 4.2 is the following discrete version of Theorem 4.1.

\proclaim{Theorem 4.3}
Let $g$ be a positive-valued even function defined on ${\Bbb R} \setminus (2\pi {\Bbb Z})$ that is periodic
with period $2\pi$. Suppose that $g$ is non-increasing and
strictly convex on $(0,\pi]$. Let $\omega_n \subset [0,2\pi)$. Let $\omega_n \subset [0,2\pi)$.
We have
$$\max_{\omega_n \in \Omega_n} \left\{ \min_{t \in \widetilde{\omega}_{2n}} {P_{\omega_n}(t)} \right\} = P_{\widetilde{\omega}_n}(\pi/(2n))\,,$$
and equality holds when $\omega_n = \omega_n^* = \{t_1^* < t_2^* < \cdots < t_n^*\}$ with
$$t_j^* = \frac{\pi}{2n} + \frac{2(j-1)\pi}{n}\,, \qquad j=1,2,\ldots, n\,.$$
\endproclaim

\demo{Proof of Theorem 4.3}
Let $\gamma$ be the number guaranteed by Theorem 4.2.
Observe that $\widetilde{\omega}_{2n}$ has exactly two points in the interval $(\gamma,\gamma + 2\pi/n)$ (mod $2\pi$).
Denote these points by $\alpha$ and $\beta = \alpha + \pi/n$. Due to the fact that $P_{\widetilde{\omega}_n}$ is
non-increasing on $(0,\pi/n)$ and
$$P_{\widetilde{\omega}_n}(t) = P_{\widetilde{\omega}_n}(2\pi/n - t)\,, \qquad t \in (0,2\pi/n)\,,$$
we have
$$\min\{P_{\widetilde{\omega}_n}(\alpha - \gamma), P_{\widetilde{\omega}_n}(\beta - \gamma)\} \leq P_{\widetilde{\omega}_n}(\pi/(2n))\,,$$
which finishes the proof of the inequality of the theorem. The fact that equality holds in the case
described in the theorem is obvious.
\qed \enddemo

Associated with $\omega_n := \{t_1 < t_2 < \cdots < t_n\} \subset [0,2\pi)$ let
$$Q_{\omega_n}(t) := \prod_{j=1}^n{\sin \left( \frac{t-t_j}{2} \right)}\,.$$
Let
$$T_n(t) := Q_{\widetilde{\omega}_n}(t) = \sin \left(\frac{nt}{2}\right)\,.$$
Our next three theorems are consequences of Theorems 4.2, and 4.3, respectively.

\proclaim{Theorem 4.4}
There is a number $\gamma \in [0,2\pi)$ (depending on $\omega_n$) such that
$$-(\log|Q_{\omega_n}|)^{(m)}(t) \leq  -(\log|T_n|)^{(m)}(t)\,, \qquad t \in (\gamma,\gamma + 2\pi/n)\,,$$
for every $\omega_n \in \Omega_n$ and for every even integer $m$.
\endproclaim

\proclaim{Theorem 4.5} Let
$$E(\omega_n) := [0,2\pi) \setminus \bigcup_{j=1}^n{ \left(t_j - \pi/n, t_j + \pi/n \right)} \quad (\text {\rm mod} \enskip 2\pi)\,.$$
We have
$$\max_{\omega_n \in \Omega_n} \left\{ \min_{t \in E(\omega_n)}{-(\log|Q_{\omega_n}|)^{(m)}(t)} \right\} = -(\log|T_n|)^{(m)}(\pi/n)$$
for every even integer $m$.
\endproclaim

\proclaim{Theorem 4.6}
We have
$$\max_{\omega_n \in \Omega_n} \left\{ \min_{t \in \widetilde{\omega}_{2n}}{-(\log|Q_{\omega_n}|)^{(m)}(t)} \right\} = -(\log|T_n|)^{(m)}(\pi/(2n))\,,$$
for every even integer $m$, and equality holds when $\omega_n = \omega_n^* = \{t_1^* < t_2^* < \cdots < t_n^*\}$ with
$$t_j^* = \frac{\pi}{2n} + \frac{2(j-1)\pi}{n}\,, \qquad j=1,2,\ldots, n\,.$$
\endproclaim

\demo {Proof of Theorem 4.4}
For the sake of brevity let $Q := Q_{\omega_n}(t)$.
Let $t \notin \omega_n \enskip (\text {\rm mod} \enskip 2\pi)$.
We have
$$(\log|Q|)^{\prime\prime}(t) = \left( \frac{Q^\prime}{Q} \right)^\prime (t) = \
\frac{d}{dt} \left( \frac 12 \sum_{j=1}^n{\cot \left( \frac{t-t_j}{2} \right)} \right)
= -\frac 14 \sum_{j=1}^n{\csc^2 \left( \frac{t-t_j}{2} \right)}\,,$$
and hence
$$-(\log|Q|)^{(m)}(t) = \frac 14 \sum_{j=1}{f^{(m-2)}(t-t_j)} = \sum_{j=1}^n{g_m(t-t_j)}\,,$$
where $f(t) := \csc^2(t/2)$ and $g_m(t) := \frac 14 f^{(m-2)}(t)$.
It is well known and elementary to check that
$$\tan t = \sum_{j=1}^n{a_jt^j}\,, \qquad t \in (-\pi/2,\pi/2)\,,$$
with each $a_j \geq 0$, $j=0,1,\ldots$. Hence, if $h(t) = \tan(t/2)$, then
$$h^{(k)}(t) > 0, \qquad t \in (0,\pi), \qquad k=0,1,\ldots\,.$$
Now observe that
$$f(t) = \csc^2 \left( \frac t2 \right) = \sec^2{\frac{\pi - t}{2}} = 2h^\prime(\pi - t)\,,$$
and hence,
$$(-1)^kf^{(k)}(t) = 2h^{(k+1)}(\pi - t) > 0, \qquad t \in (0,\pi)\,.$$
This implies that if $m$ is even $g_m(t) = \frac 14 f^{(m-2)}(t)$ is a positive, decreasing, strictly convex function on $(0,\pi)$.
It is also clear that if $m$ is even, then $g_m$ is even since $f$ is even.
Now we can apply Theorem 4.2 to deduce that there is a number $\gamma \in [0,2\pi)$ (depending on $\omega_n$) such that
$$-(\log|Q_{\omega_n}|)^{(m)}(t) = \sum_{j=1}^n{g_m(t-t_j)} \leq -(\log|T_n|)^{(m)}(t)\,, \qquad t \in [\gamma,\gamma + 2\pi/n)\,,$$
and the proof is finished.
\qed \enddemo

\demo {Proof of Theorem 4.5}
The theorem follows from Theorem 4.4 immediately.
\qed \enddemo

\demo {Proof of Theorem 4.6}
We use the notation and the observations in the proof of Theorem 4.4. However, at the end of the proof we use Theorem 4.3 to deduce that
$$\min_{t \in \widetilde{\omega}_{2n}}Q_{\omega_n}(t) \leq T_n(\pi/(2n))\,,$$
and equality holds when $Q_{\omega_n} = T_n$.
\qed \enddemo

We conclude this section by giving an independent proof of the unit circle polarization conjecture in [ABE-12] for the
case $p=4$, where we show that, for $z_1, z_2, \ldots, z_n \in {\Bbb S}^1,$ a ``good polarization point"
$z_0 \in {\Bbb S}^1$ can be chosen so that
$$\prod_{j=1}^n{|z_0-z_j|} = \max_{z \in {\Bbb S}^1}{\prod_{j=1}^n{|z-z_j|}}\,. \tag 4.1$$

\proclaim{Theorem 4.7}
If $z_1, z_2, \ldots, z_n \in {\Bbb S}^1$,
then
$$\min_{z \in {\Bbb S}^1}{\sum_{j=1}^n{\frac{1}{|z-z_j|^4}}} \leq
\frac{n^4}{48} + \frac{n^2}{24}, \qquad\,\, n \geq 1,$$
and equality holds when the points $z_j$ are distinct and equally spaced on ${\Bbb S}^1;$
that is, (1.5) holds. Moreover, if $z_1, z_2, \ldots, z_n \in {\Bbb S}^1$, and $z_0 \in {\Bbb S}^1$ is chosen
so that (4.1) holds, then
$$\sum_{j=1}^n{\frac{1}{|z_0-z_j|^4}} \leq \frac{n^4}{48} + \frac{n^2}{24}, \qquad\,\, n \geq 1.$$
\endproclaim

This result naturally suggests the following open question:

\proclaim{Problem}
For what values of $p \in (0,\infty)$ is it true that
$$\sum_{j=1}^n{\frac{1}{|z_0-z_j|^p}} \leq M_n^p({\Bbb S}_1)$$
whenever $z_1,z_2,\ldots,z_n \in {\Bbb S}_1$ and $z_0 \in {\Bbb S}_1$ satisfies (4.1) ?
\endproclaim

In addition to the value $p=4$,  a closer look at the main result in [ABE-12] shows that $p = 2$ is also such a value.

\demo{Proof of Theorem 4.7}
Write $z_j=e^{it_j},\,t_j \in [0,2\pi),\, j=1,2,...,n,$ and
set
$$Q_n(t) := \prod_{j=1}^{n}{\sin{\frac{t-t_j}{2}}}\,.$$
Then $H_n$ defined by $H_n(t) := Q_n(2t)$ is a real trigonometric polynomial
of degree $n$.
We have the following identities:
$$\frac{Q_n^{\prime}(t)}{Q_n(t)} = \frac 12 \sum_{j=1}^{n}{\cot{\frac{t-t_j}{2}}}\,,$$
$$\left( \frac{Q_n^{\prime}}{Q_n} \right)^\prime(t) =
-\frac 14\sum_{j=1}^{n}{\csc^2{\frac{t-t_j}{2}}} =
-\frac 14 \sum_{j=1}^{n}{\sin^{-2}{\frac{t-t_j}{2}}}\,,$$
$$\left( \frac{Q_n^{\prime}}{Q_n} \right)^{\prime\prime}(t) =
-\frac 14 \sum_{j=1}^{n}{\frac 12 \cos{\frac{t-t_j}{2}} (-2) \sin^{-3}{\frac{t-t_j}{2}}}
= \frac 14 \sum_{j=1}^{n}{\cos{\frac{t-t_j}{2}} \sin^{-3}{\frac{t-t_j}{2}}}\,,$$
$$ \left( \frac{Q_n^{\prime}}{Q_n} \right)^{\prime\prime\prime}(t) = \frac 14 \sum_{j=1}^{n}
{\left(\sin^{-2}{\frac{t-t_j}{2}} - \frac 32 \sin^{-4}{\frac{t-t_j}{2}} \right)}\,,$$
so
$$\frac 38 \sum_{j=1}^{n}{\sin^{-4}{\frac{t-t_j}{2}}} =
-\left( \frac{Q_n^{\prime}}{Q_n} \right)^{\prime\prime\prime}(t) -
\left( \frac{Q_n^{\prime}}{Q_n} \right)^\prime(t)\,.$$
On the other hand,
$$\left( \frac{Q_n^{\prime}}{Q_n} \right)^{\prime\prime\prime} =
\frac{Q_n^{(4)}}{Q_n} - 3Q_n^{\prime\prime\prime}\frac{Q_n^\prime}{Q_n^2} -
3Q_n^{\prime\prime}\left(\frac{Q_n^{\prime\prime}Q_n^2 -
2Q_nQ_n^\prime Q_n^\prime}{Q_n^4}\right) +
Q_n^\prime \left( \frac{1}{Q_n}\right)^{\prime\prime\prime}$$
and
$$\left( \frac{Q_n^{\prime}}{Q_n} \right)^\prime =
\frac{Q_n^{\prime\prime}}{Q_n} - \left(\frac{Q_n^\prime}{Q_n}\right)^2\,.$$
Hence
$$\left( \frac{Q_n^{\prime}}{Q_n} \right)^{\prime\prime\prime}(t_0) =
\frac{Q_n^{(4)}}{Q_n}(t_0) - 3 \left(\frac{Q_n^{\prime\prime}}{Q_n} \right)^2(t_0)$$
and
$$\left( \frac{Q_n^{\prime}}{Q_n} \right)^\prime (t_0)= \frac{Q_n^{\prime\prime}}{Q_n}(t_0)$$
at every point $t_0$ such that $Q_n^\prime(t_0) = 0$. So if
$z_0 = e^{it_0} \in {\Bbb S}^1$ is chosen so that
$$|Q_n(t_0)| = \max_{t \in [-\pi,\pi]}{|Q_n(t)|}\,,$$
then
$$\split 6\sum_{j=1}^{n}{\frac{1}{|z_0-z_j|^4}} = &
\left( 3\left( \frac{Q_n^{\prime\prime}}{Q_n} \right)^2
- \frac{Q_n^{(4)}}{Q_n} - \frac{Q_n^{\prime\prime}}{Q_n} \right)(t_0) \cr = &
\left( \frac{3}{16}\left( \frac{H_n^{\prime\prime}}{H_n} \right)^2
- \frac{1}{16}\frac{H_n^{(4)}}{H_n} - \frac{1}{4}\frac{H_n^{\prime\prime}}{H_n} \right)
\left(\frac{t_0}{2}\right)\,.
\cr \endsplit$$
Without loss of generality we may assume that $t_0 = 0$ and $z_0 = 1$.

Set
$$F(H_n) := \left(\frac{3}{16}(H_n^{\prime\prime})^2 - \frac{1}{16}H_n^{(4)}
- \frac{1}{4}H_n^{\prime\prime}\right)(0)\,$$
and let ${\Cal A}_n$ be the set of all real trigonometric polynomials
$H_n$ of degree at most $n$ such that
$$H_n(0) = \max_{t \in [-\pi,\pi]}{|H_n(t)|} = 1\,.$$
A simple compactness argument shows that there is a $\widetilde{H}_n \in {\Cal A}_n$ such that
$$F(\widetilde{H}_n) = \sup_{H_n \in {\Cal A}_n}{F(H_n)}\,.$$
Let
$$\widetilde{U}_n(t) := \frac 12 (\widetilde{H}_n(t) + \widetilde{H}_n(-t))\,.$$
Then $\widetilde{U}_n \in {\Cal A}_n$ is even and $F(\widetilde{U}_n) = F(\widetilde{H}_n)$.
Since $\widetilde{U}_n \in {\Cal A}_n$ is even, it is of the form
$$\widetilde{U}_n(t) =: \widetilde{P}_n(\cos t)$$
for a $\widetilde{P}_n \in {\Cal P}_n$ satisfying
$$\widetilde{P}_n(1) = \max_{x \in [-1,1]}{|\widetilde{P}_n(x)|} = 1\,,$$
where ${\Cal P}_n$ denotes the set of all real algebraic polynomials of degree at most $n$.

Observe that $U_n \in {\Cal A}_n$ is even if and only if it is of the form
$$U_n(t) =: P_n(\cos t)$$
for a $P_n \in {\Cal P}_n$ satisfying
$$P_n(1) = \max_{x \in [-1,1]}{|P_n(x)|} = 1\,.$$
A simple calculation shows that
$$U_n(0)  = P_n(1),\quad
U_n^{\prime\prime}(0) = -P_n^\prime(1),\quad
U_n^{(4)}(0) = 3P_n^{\prime\prime}(1) + P_n^\prime(1)\,. $$
Let
$$\split G(P_n) := & F(U_n) = \left(\frac{3}{16}(U_n^{\prime\prime})^2
- \frac{1}{16}U_n^{(4)} - \frac{1}{4}U_n^{\prime\prime}\right)(0) \cr
= & \frac{3}{16}((P_n^\prime)^2 - P_n^{\prime\prime} + P_n^{\prime})(1)\,. \cr \endsplit$$
We have
$$G(P_n) = F(U_n) \leq F(\widetilde{H}_n) = F(\widetilde{U}_n) = G(\widetilde{P}_n)$$
for every $P_n \in {\Cal P}_n$ such that
$$P_n(1) = \max_{x \in [-1,1]}{|P_n(x)|} = 1\,.$$\

Next we show by a simple variational method that $\widetilde{P}_n \in {\Cal P}_n$
equioscillates between $-1$ and $1$ at least $n$ times on $[-1,1]$. That is, there are
$$-1 \leq y_n < y_{n-1} < \cdots < y_1 = 1$$
such that
$$\widetilde{P}_n(y_j) = (-1)^{j-1}\,, \qquad j=1,2,\ldots,n\,.$$  To show this, first we observe that $\widetilde{P}_n^\prime(1) > 0$ since
$\widetilde{P}_n^\prime(1) \geq 0$, and Markov's inequality for the second
derivative (see p. 249 of [BE-95]) together with $\widetilde{P}_n^\prime(1) = 0$
would imply that
$$\split G(\widetilde{P}_n) & =
\frac{3}{16}((\widetilde{P}_n^\prime)^2 - \widetilde{P}_n^{\prime\prime} +
\widetilde{P}_n^{\prime})(1)
= \frac{-3}{16}\widetilde{P}_n^{\prime\prime}(1) \cr
& \leq \frac{3}{16}T_n^{\prime\prime}(1) =
\frac{3}{16} \, \frac{n^2(n^2-1)}{3} <
\frac{1}{16}(2n^4 + 4n^2) = G(T_n)\,, \cr \endsplit$$
where $T_n$ is the Chebyshev polynomial of degree $n$ defined by
$T_n(\cos t) = \cos(nt)$, and this contradicts the extremal property of $\widetilde{P}_n$.
Now let
$$E := \{y \in [-1,1]: |\widetilde{P}(y)|=1\}\,.$$
We list the elements of $E$ as
$$E = \{1=y_1 > y_2 > \cdots > y_{\mu}\}\,,$$
where
$$\widetilde{P}_n(y_{k_j}) = \widetilde{P}_n(y_{k_j+1}) = \cdots
= \widetilde{P}_n(y_{k_{j+1}-1})\,, \qquad j=0,1,\ldots,m-1\,,$$
and
$$\widetilde{P}_n(y_{k_j}) = - \widetilde{P}_n(y_{k_j-1}) = (-1)^j\,,
\qquad j=1,2,\ldots,m-1\,,$$
for some
$$1 = k_0 < k_1 < \cdots < k_m = \mu+1\,.$$
Now we pick
$$\alpha_j \in (y_{k_j},y_{k_j-1})\,, \qquad j=1,2,\ldots,m-1\,.$$

Assume that $m \leq n-1$. For the polynomial $R_n \in {\Cal P}_n$ defined by
$$R_n(x) := (x-1)^2\prod_{j=1}^{m-1}{(x-\alpha_j)}\,$$
we have
$$R_n(y)\widetilde{P}_n(y) > 0\,, \qquad y \in E \setminus \{1\}\,,$$
$$R_n(1) = R_n^{\prime}(1) = 0 \qquad \text {and} \qquad R_n^{\prime\prime}(1) > 0\,.$$
These properties together with $\widetilde{P_n}^\prime(1) > 0$ imply that
for a sufficiently small value of $\varepsilon > 0$ the polynomial
$$S_n = \widetilde{P}_n - \varepsilon R_n \in {\Cal P}_n$$
satisfies
$$S_n(1) = \max_{x \in [-1,1]}{|S_n(x)|} = 1$$
and $G(S_n) > G(\widetilde{P}_n)$, so $S_n \in {\Cal P}_n$ contradicts the extremal
property of $\widetilde{P}_n$. This finishes the proof of the fact that
$\widetilde{P}_n \in {\Cal P}_n$ equioscillates between $-1$ and $1$ at least $n$ times
on $[-1,1]$, as we claimed.

As a consequence, the Intermediate Value Theorem implies that
$\widetilde{P}_n$ has at least $n-1$ zeros in $(-1,1)$, say
$$(-1 <) x_{n-1} < x_{n-2} < \cdots < x_1 (< 1)\,.$$
Observe that the polynomial $\widetilde{P}_n \in {\Cal P}_n$ has
an odd number of zeros (by counting multiplicities) in each of the
intervals $(y_{j+1},y_j)$ for $j=1,2,\ldots,n-1$; hence $x_j$ is
the only (simple) zero of $\widetilde{P}_n$ in $(y_{j+1},y_j)$
for each $j=1,2,\ldots,n-1$.
Therefore $\widetilde{P}_n$ has only real zeros and it is of the form
$$\widetilde{P}_n(x) = c\prod_{j=1}^{\mu}{(x-x_j)}$$
with either $\mu=n-1$ or $\mu=n$, and in the case $\mu=n$ we have
$x_n \in {\Bbb R} \setminus [y_n,1]$.

 Note that
$$\frac{\widetilde{P}_n^\prime(x)}{\widetilde{P}_n(x)} = \sum_{j=1}^\mu{\frac{1}{x-x_j}}, \quad
\left( \frac{\widetilde{P}_n^\prime(x)}{\widetilde{P}_n(x)} \right)^\prime =
-\sum_{j=1}^\mu{\frac{1}{(x-x_j)^2}}\,,$$
and
$$G(\widetilde{P}_n) =
\frac{3}{16}\,\left( \frac{(\widetilde{P}_n^\prime)^2 -
\widetilde{P}_n^{\prime\prime}\widetilde{P}_n}{(\widetilde{P}_n)^2}
+ {\widetilde{P}_n^{\prime}}{\widetilde{P}_n}\right)(1)
= \frac{3}{16} \,\left( \sum_{j=1}^\mu{\frac{1}{(1-x_j)^2}} +
\sum_{j=1}^\mu{\frac{1}{(1-x_j)}}\right)\,.$$
If $\mu = n-1$, then $\widetilde{P}_n$ equioscillates between $-1$ and $1$ on $[-1,1]$
the maximum number of times, so $\widetilde{P}_n \equiv T_{n-1}$, where $T_{n-1}$
is the Chebyshev polynomial of degree $n-1$ defined by $T_{n-1}(\cos t) = \cos((n-1)t)$.
Hence
$$\split G(\widetilde{P}_n) & =
\frac{3}{16} \,\left( \sum_{j=1}^\mu{\frac{1}{(1-x_j)^2}} +
\sum_{j=1}^\mu{\frac{1}{(1-x_j)}}\right) \cr
& = \frac{3}{16}\,\left( \frac{(T_{n-1}^\prime)^2 -
T_{n-1}^{\prime\prime}T_{n-1}}{T_{n-1}^2}
+ {T_{n-1}^{\prime}}{T_{n-1}}\right)(1) \cr
& = \frac{3}{16}\left((n-1)^4 - \, \frac{(n-1)^2((n-1)^2-1)}{3} + (n-1)^2 \right)
= \frac 18 (n-1)^4 + \frac 14 (n-1)^2\,. \cr \endsplit$$
If $\mu = n$ we must have
$x_n \in (-\infty,y_n) \cup (1,\infty)$. However, $1 < x_n$ would imply that
$$Y_n(x) := -c(x-(2-x_n))\prod_{j=1}^{n-1}{(x-x_j)}$$
satisfies
$$Y_n(1) = \max_{x \in [-1,1]}{|Y_n(x)|} = 1$$
and
$$G(Y_n) = G(\widetilde{P}_n)\,,$$
and hence $Y_n \in {\Cal P}_n$ also shares the extremal property of $\widetilde{P}_n$
while it has all its zeros in $(-\infty, 1)$. Hence $x_n < y_n < x_{n-1}$. But then
$\widetilde{P}_n$ is just the Chebyshev polynomial $T_n$ transformed linearly from
the interval $[-1,1]$ to $[\eta,1]$ for some $\eta \leq -1$. This implies that
$$\split G(\widetilde{P}_n) & =
\frac{3}{16} \,\left( \sum_{j=1}^\mu{\frac{1}{(1-x_j)^2}} +
\sum_{j=1}^\mu{\frac{1}{(1-x_j)}}\right) \cr
& = \frac{3}{16}\,\left(\left( \frac{2}{1-\eta} \right)^2
\frac{(T_n^\prime)^2 - T_n^{\prime\prime}T_n}{T_n^2}
+ \frac{2}{1-\eta} {T_n^{\prime}}{T_n}\right)(1) \cr
&\leq \frac{3}{16}\left(n^4 - \, \frac{n^2(n^2-1)}{3} + n^2 \right)
= \frac 18 n^4 + \frac 14 n^2\,. \cr \endsplit$$
Now we conclude that
$$G(\widetilde{P}_n) \leq G(T_n) = \frac 18 n^4 + \frac 14 n^2\,,$$
and hence
$$F(\widetilde{H}_n) = G(\widetilde{P}_n) \leq G(T_n) = \frac 18 n^4 + \frac 14 n^2\,.$$
Therefore
$$6\sum_{j=1}^{n}{\frac{1}{|z_0-z_j|^4}} = F(H_n) \leq  F(\widetilde{H}_n)
\leq G(T_n) = \frac 18 n^4 + \frac 14 n^2\,,$$
and this completes the proof.
\qed \enddemo

We conclude this section by mentioning two formulas that may be useful for future
investigations of the polarization problem for the unit circle. Let
$$A_p(t) := \sum_{j=1}^n{\frac{1}{|e^{it}-z_j|^p}}\,, \qquad p > 0\,,$$
where  $z_j = e^{it_j} \in {\Bbb S}^1, j=1,2,\ldots, n$. 
Then straightforward calculation yields the following:
$$A_2(t) = \frac{Q_n^{\prime\prime}(t)Q_n(t) - (Q_n^\prime(t))^2}{(Q_n(t))^2} \qquad \text {\rm with} \qquad Q_n(t) := \prod_{j=1}^n{\sin \frac{t-t_j}{2}}\,,$$
and
$$A_{p+2}(t) = \frac{1}{p^2+p}(A_p^{\prime\prime}(t) + p^2A_p(t)).$$

\head 5. Proofs of Theorems 2.4, 3.1, and 3.2  \endhead

\demo{Proof of Theorem 2.4}
We proceed with an argument similar to that in [KS-98]. Let
$\omega_n= \{{\bold x}_j\}_{j=1}^n \subset A.$  Setting
$$r_n:=(2nC_0)^{-1/d},\quad D_j:= A \setminus B^m({\bold x}_j,r_n), \quad
{\Cal D}:=\cap_{j=1}^n D_j,$$
we have from (2.14) that
$$\mu({\Cal D}) \geq 1-\sum_{j=1}^n\mu(B^m({\bold x}_j,r_n)\cap A) \geq
1-nC_0r_n^d=\frac{1}{2}.$$
Thus, for
$$f_n({\bold x}):=\sum_{j=1}^n|{\bold x} - {\bold x}_j|^{-p},$$
we obtain
$$M^p(\omega_n,A)\leq \frac{1}{\mu({\Cal D})}\int_{{\Cal D}}
{f_n({\bold x})\, d\mu({\bold x})} \leq
2\sum_{j=1}^n\int_{D_j}|{\bold x} - {\bold x}_j|^{-p}\,d\mu({\bold x}). \tag 5.1$$
Next, we bound the integrals over $D_j$ utilizing (2.14):
$$\split \int_{D_j}{|{\bold x} - {\bold x}_j|^{-p}\,d\mu({\bold x})} & =
\int_0^{\infty}{\mu\{{\bold x}\in D_j:|{\bold x} - {\bold x}_j|^{-p}>t\}\, dt} \cr
& \leq 1 + \int_1^{r_n^{-p}}{\mu(B^m({\bold x}_j,t^{-1/p}) \cap A)\,dt} \cr
& \leq 1+C_0\int_1^{r_n^{-p}}{\frac{1}{t^{d/p}}\,dt}, \cr \endsplit $$
where we assume that $n$ is sufficiently large so that $r_n^{-p}>1.$
Thus  from (5.1) it follows that
$$M^p(\omega_n,A)\leq 2 n\left(1+C_0\int_1^{r_n^{-p}}\frac{1}{t^{d/p}}\,dt\right).
\tag 5.2$$
Consequently, for $p>d$ we get
$$M^p(\omega_n,A)\leq 2 n\left(1+C_0\frac{p}{p-d}[r_n^{d-p}-1]\right) \leq
\frac{c_p}{p-d}\,n^{p/d} \tag 5.3$$
and for $p=d$ we obtain
$$M^d
(\omega_n,A)\leq 2 n[1+C_0\log(r_n^{-d})]=2n[1+C_0\log(2nC_0)]\leq c_1n\log n.
\tag 5.4$$
This completes the proof of parts (2.11) and (2.12) of Theorem 2.4, while
(2.13) follows immediately upon integration of $f_n({\bold x})$ with respect to $d\mu_A.$
\qed \enddemo

\demo{Proof of Theorem 3.1}
Inequality (3.4) is an immediate consequence of (2.8), while equation (3.6) follows from
(3.2), (2.10), and the last assertion in Theorem 2.4 , since
$$\int |{\bold x} - {\bold y}|^{-p}\,d\sigma_d(y)=W_p({\Bbb S}^d),
\quad {\bold x}\in {\Bbb S}^d,\quad p<d.
\tag 5.5$$
To prove equation (3.5), we first note that from (2.9) we have
$$\liminf_{n \to \infty}\frac {M_n^p({\Bbb S}^d)}{n\log n} \geq
\frac {\beta_d}{{\Cal H}_d({\Bbb S}^d)} =
\frac 1d \frac{\Gamma \big( \frac{d+1}{2} \big)}
{\sqrt{\pi} \, \Gamma \big( \frac d2 \big)} = \tau_d.$$
Hence, if we establish the upper estimate in (3.7) for $p=d,$ then (3.5) will follow.
For this purpose, we refine the argument used in the proof of Theorem 2.4. With
$\mu=\sigma_d,$ the following estimates are known for
${\bold x} \in {\Bbb S}^d $
(cf. [KS-98]):
$$\sigma_d(B^{d+1}({\bold x},r) \cap{\Bbb S}^d) \leq \tau_dr^d, \tag 5.6$$
and
$$ \int_{{\Bbb S}^d \setminus B^{d+1}({\bold x},r)}|{\bold x} - {\bold y}|^{-d}\,d\sigma_d({\bold y} = d\tau_d2^{-d/2}\int_{-1}^{1-r^2/2}(1-t)^{-1}(1+t)^{\frac{d}{2}-1}\, dt
\leq d\tau_d \log(2/r)\,,$$
for $0<r<2.$ Utilizing these estimates and using (5.1) with
$r_n=(\tau_dn\log n)^{-1/d},$
$D_j={\Bbb S}^d \setminus B^{d+1}({\bold x}_j,r_n)$, and $n \geq 3$, we obtain
$$M^d(\omega_n,A)\leq \frac{1}{1-n\tau_dr_n^d}\sum_{j=1}^n\int_{D_j}
{|{\bold x} - {\bold x}_j|^{-d}\,d\sigma_d({\bold x})} \leq
\frac{nd}{1-n\tau_dr_n^d}\tau_d\log(2/r_n)$$
$$ =\frac{nd}{1-(\log n)^{-1}}\tau_d \left( \log 2 +\frac{1}{d}\log(\tau_d n \log n) \right)$$
$$=\tau_d\frac{n[\log n +\log(\log n)+\log(2^d\tau_d) ]}{1-(\log n)^{-1}}.$$
This completes the proof of (3.5) as well as the upper bound in (3.7) for the case $p=d.$

It remains to establish (3.7) for the cases $p<d$ and $p>d.$ But, as observed above, the
former is a consequence of (2.13) and (5.5). So hereafter we assume $p>d.$
From the estimate
$$ \int_{{\Bbb S}^d \setminus B^{d+1}({\bold x},r)}
{|{\bold x} - {\bold y}|^{-p}\,d\sigma_d({\bold y})}
= d\tau_d2^{-p/2}\int_{-1}^{1-r^2/2}
{(1-t)^{-\frac{p}{2}+\frac{d}{2}-1}(1+t)^{\frac{d}{2}-1}\, dt}$$
$$ \leq d\tau_d2^{-\frac{p}{2}+\frac{d}{2}-1}
\int_{-1}^{1-r^2/2}(1-t)^{-\frac{p}{2}+\frac{d}{2}-1}\, dt
= \frac{d\tau_d}{p-d}[r^{-p+d}-2^{-p+d}], \quad r<2\,,$$
and inequality (5.5), we deduce (as above) that
$$M^p(\omega_n,A) \leq \frac{n}{1-n\tau_d r^d}\left(\frac{d\tau_d}{p-d}\right)r^{-p+d}.
\tag 5.7$$
In this case, an optimal choice for $r$ is
$$r=r_n=\left(\frac{p-d}{np\tau_d}\right)^{1/d},$$
which when substituted in (5.7) yields the estimate stated in inequality (3.7) for the
case $p>d.$
\qed \enddemo

\demo{Proof of Theorem 3.2} Assertion (3.8) is immediate from (2.8). Also
the upper bounds in (3.12) for the cases $p>d$ and $p=d,$
can be established in same way as in the proof of Theorem 3.1, with the measure
$\sigma_d$ replaced by normalized $d$-dimensional Lebesgue measure (volume measure). We leave
the details for the reader. Furthermore, (3.9) follows from (3.12) together with Theorem 2.2.

 For the case $d-2<p<d, p>0$, the upper estimate in (3.12) follows from (3.3),
(2.13), and the fact that
$$\int\frac{1}{|{\bold x}-{\bold y}|^p}\,d\mu_p({\bold y}) \leq
W_p({\Bbb B}^d),\quad \quad {\bold x} \in {\Bbb B}^d,$$
where $\mu_p$ is the $p$-equilibrium probability measure on ${\Bbb B}^d$ (cf. [L-72]).
Together with (2.10), we also deduce (3.11). (Alternatively,
one can apply the result of [FN-08, Theorem 11] mentioned in Section 2 to deduce (3.11).)

It remains to establish (3.10). For this purpose observe that for the range $0<p<d-2$, the kernel
$K({\bold x},{\bold y})=|{\bold x}-{\bold y}|^{-p}$ is superharmonic, so that the minimum principle applies.
Let $\omega_n=\{{\bold x}_1,{\bold x}_2, \ldots, {\bold x}_n\}$ be a list of $n$ points (not necessarily distinct) in ${\Bbb B}^d$
and set
$$U({\bold x}):=\sum_{k=1}^{n}\frac{1}{|{\bold x}-{\bold x}_k|^p}.$$
We claim that
$$M^p(\omega_n,{\Bbb B}^d)=\min\{U({\bold x}): {\bold x} \in {\Bbb B}^d \} \leq n,\tag 5.8$$
from which (3.10) will follow, since on taking all points ${\bold x}_k$  to be at zero, we
get that $M_n^p({\Bbb B}^d) \geq n.$  To establish (5.8), let $\sigma_{d-1}$ denote normalized surface area
measure on the boundary ${\Bbb S}^{d-1}$ of ${\Bbb B}^d.$  By the minimum principle we have
$$M^p(\omega_n,{\Bbb B}^d)=\min\{U({\bold x}): {\bold x} \in {\Bbb S}^{d-1} \} \leq \int_{{\Bbb S}^{d-1}} U({\bold x})\, d\sigma_{d-1}({\bold x}).\tag 5.9$$
Again applying the minimum principle,  it follows that
$$V({\bold y}):= \int_{{\Bbb S}^{d-1}}\frac{1}{|{\bold x}-{\bold y}|^p} \, d\sigma_{d-1}({\bold x})$$
satisfies $1=V({\bold 0}) \geq \min\{V({\bold y}): |{\bold y}|=r \}$ for each $0 \leq r \leq 1.$ But
as is easily seen, $V$ is constant on each sphere $|{\bold y}|=r,$ from which we deduce that $1 \geq V({\bold y})$ for
all ${\bold y} \in {\Bbb B}^d.$  Therefore, from (5.9) we obtain
$$M^p(\omega_n,{\Bbb B}^d) \leq \sum_{k=1}^n V({\bold x}_k) \leq n, $$
which establishes the claim and completes the proof.

\qed \enddemo

\noindent ${\bold {Acknowledgment}}$ The authors are grateful to Doug Hensley for his
initial involvement and insight on the topic of this manuscript.

\enddocument